\documentclass[twoside,onecolumn]{article}
\usepackage{blindtext} 
\usepackage{graphicx}
\usepackage[sc]{mathpazo} 
\usepackage[T1]{fontenc} 
\usepackage{microtype} 
\usepackage[english]{babel} 
\usepackage[hmarginratio=1:1,top=32mm,columnsep=20pt]{geometry} 
\usepackage[hang, small,labelfont=bf,up,textfont=it,up]{caption} 

\usepackage{booktabs} 
\usepackage{lettrine} 
\usepackage{enumitem} 
\setlist[itemize]{noitemsep} 
\usepackage{abstract} 
\usepackage{titlesec} 
\renewcommand\thesection{\Roman{section}} 
\renewcommand\thesubsection{\roman{subsection}} 

\newcommand\ffbox[1]{#1}
\usepackage{amsmath}

\usepackage{tikz}
\usepackage{tikzscale}

\titleformat{\section}[block]{\large\scshape\centering}{\thesection.}{1em}{} 
\titleformat{\subsection}[block]{\large}{\thesubsection.}{1em}{} 
\usepackage{fancyhdr} 
\usepackage{titling} 
\usepackage{hyperref} 
\pretitle{\begin{center}\Large\bfseries} 
\posttitle{\end{center}} 
\title{Richtmyer-Meshkov mixing layer growth from localized perturbations} 
\author{%
\textsc{Britton J. Olson} \\ 
\normalsize Lawrence Livermore National Laboratory, Livermore, CA, USA \\ 
\normalsize olson45@llnl.gov \\ 
\textsc{Robin Williams} \\ 
\normalsize Atomic Weapons Establishment, United Kingdom  
}
\date{} 
\renewcommand{\maketitlehookd}{%
\begin{abstract}
\noindent

We study the growth of Richtmyer-Meshkov \cite{richtmyer.1960,meshkov.1969} mixing layers from an initial surface with spatially localized perturbations.  We use two symmetric forms of the initial patch, which allow simulation data to be averaged to generate a two-dimensional statistical representation of the three dimensional turbulent flow.  We find that as the mixing layer grows, the turbulent structures tend to form into discrete packets separated from the surface, with material entrainment into them dominated by a laminar entrainment flow inward from the surrounding regions where the surface was originally smooth.  The entrainment appears to be controlled by the propagation of vortex pairs which appear at the boundary of the region of initial perturbations.  This suggests that the growth of RM mixing from isolated features, as may be found in manufactured Inertial Confinement Fusion capsules, has a rather different mechanism than the growth of an RM mixing layer when the perturbations are uniform.  This may be a challenge for some existing engineering models.

\end{abstract}
}
\begin{document}
%
\maketitle
%
\section{Introduction}
\label{sec:intro}
\noindent Applications in engineering and science where hydrodynamic instability at an interface leads to large scale mixing often occur in non-ideal configurations.  Many rigorous studies of Richtmyer-Meshkov (RM) and Rayleigh-Taylor (RT) (see \cite{zhou.pr.2017-1,zhou.pr.2017-2} and the references therein) exist which assume instability growth on an interface which is statistically homogeneous in all directions.  Therefore, the mean flow is inherently 1D, where the other dimensions have been collapsed, as in the case of planar and spherical instability growth.  However, there are relatively few studies on interfacial instability growth where the underlying mean flow is truly multi-dimensional.  Engineering applications such as ICF fill tubes and tent perturbations \cite{weber.pop.2017} represent mean geometries which are 3D and 2D respectively and create a challenge for engineering codes which utilizes Reynolds Averaged Navier-Stokes (RANS) closure models to represent the instability and turbulence as a subgrid scale transport model.

In these proceedings, we propose a modification to a simple RM test case which generates 2D mean flow features.  This is accomplished by localizing a patch of initial perturbations on a plane (like many other studies) but which now contain an edge or a boundary.  The lack of spatial homogeneity in these regions is representative of the aforementioned applications.  It highlights potential modeling deficiencies in current RANS engineering approaches.

In Section \ref{sec:methods} the Large-Eddy Simulation (LES) methodology is described.  In Section \ref{sec:problem} a modification to a standard planar RM test case is proposed which makes the mean flow multi-dimensional.  High fidelity LES results over a range of grid resolutions are generated to establish a bound on mesh dependence.  Two sets of initial perturbations are explored; one which resembles a strip leads to a ``curtain'' of mixing and one which resembles a patch leads to a ``plume''.

In Section \ref{sec:results} quantitative results are given comparing the vertical and horizontal mixing layer growth as a function of time.  Higher order turbulence statistics such as TKE and flow anisotropy for the different resolutions and configurations are compared, as well.
Finally, in Section \ref{sec:discussion} we summarize the present findings and discuss briefly modeling considerations using the RANS modeling approach and conclude.

%
%
\section{Methods}
\label{sec:methods}
\noindent Large Eddy Simulations (LES) of the three dimensional Navier-Stokes equations are solved using the Ares code developed at Lawrence Livermore National Laboratory.  The full equations of motion and the numerical methods solved in the Ares code are detailed in \cite{olson.pof.2014} and \cite{sharp.llnl.1981}.  For the present calculations, the infinite Reynolds number limit is assumed as in \cite{thornber.pof.2017} and no physical transport properties are used (viscosity, conductivity, diffusivity, etc.).  Therefore, three grid resolutions are explored to quantify the grid resolution dependence on the quantities of interest.
The coarse, medium, and fine mesh resolutions are given as  (128$\times$128$\times$192), (256$\times$256$\times$384), and (512$\times$512$\times$768) grid points, respectively, with a domain size is given as $2.8\pi\times 2\pi \times 2\pi$.


%
\section{Localized perturbation problem description}
\label{sec:problem}
\noindent We present here a numerical experiment setup which can be used to generate non one-dimensional mean flow, unlike that typically assumed for RM mixing layers.  To do this, we use the work of Thornber et al \cite{thornber.pof.2017} as the initial conditions subject to modifications to localize the perturbations into patches and create edges to the mixing layer.  A Mach 1.84 shock drives the mixing growth as it traverses from heavy ($\rho=3$) to light ($\rho=1$) fluids, both with ideal gases with $\gamma=5/3$.

\begin{figure}[h]
\centering
\includegraphics[width=.3\textwidth]{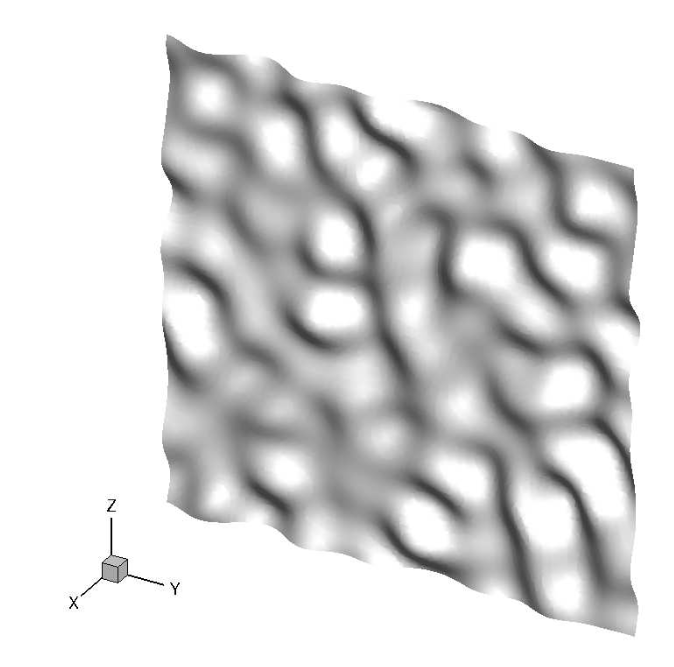}
\includegraphics[width=.3\textwidth]{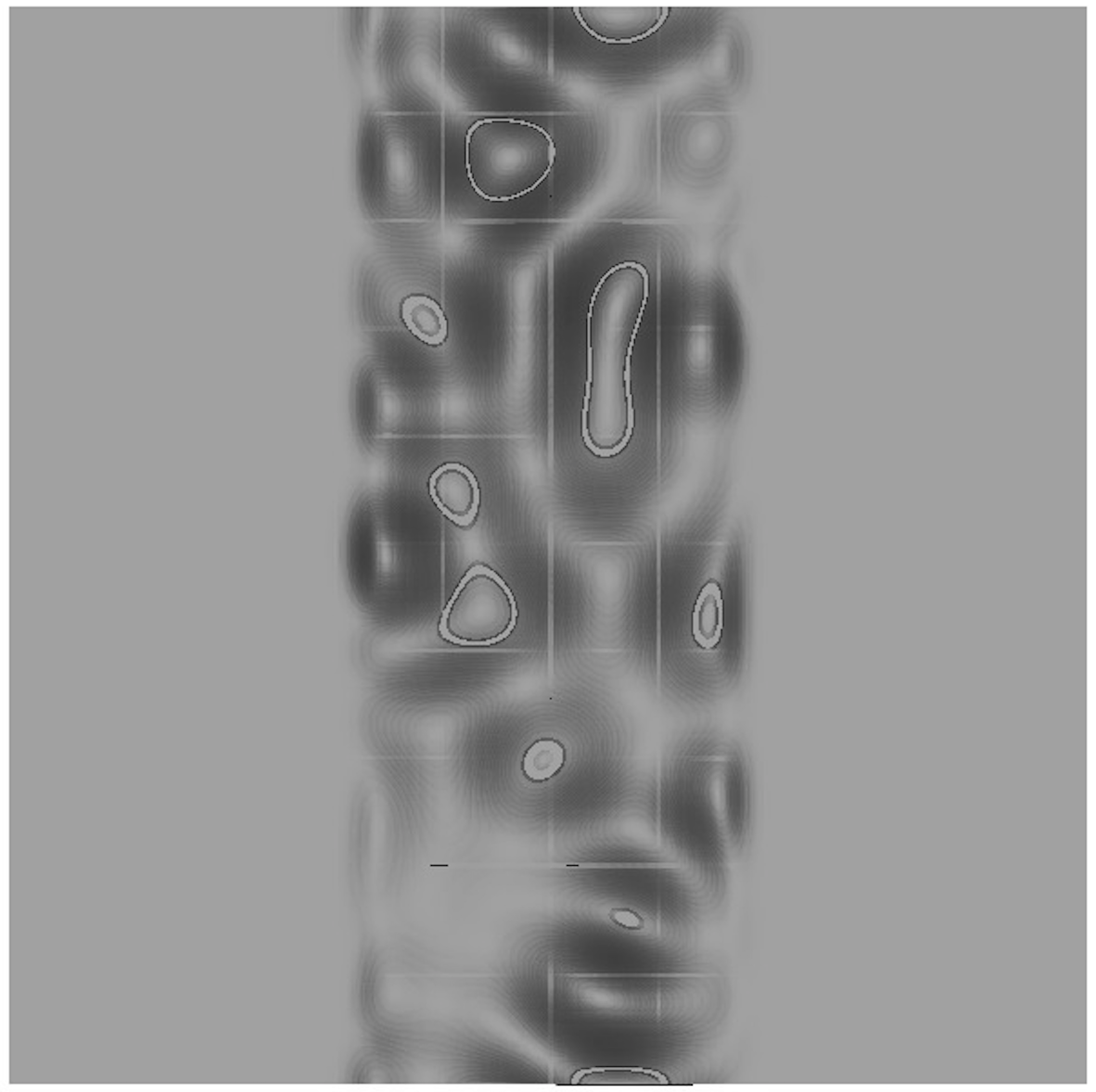}
\includegraphics[width=.3\textwidth]{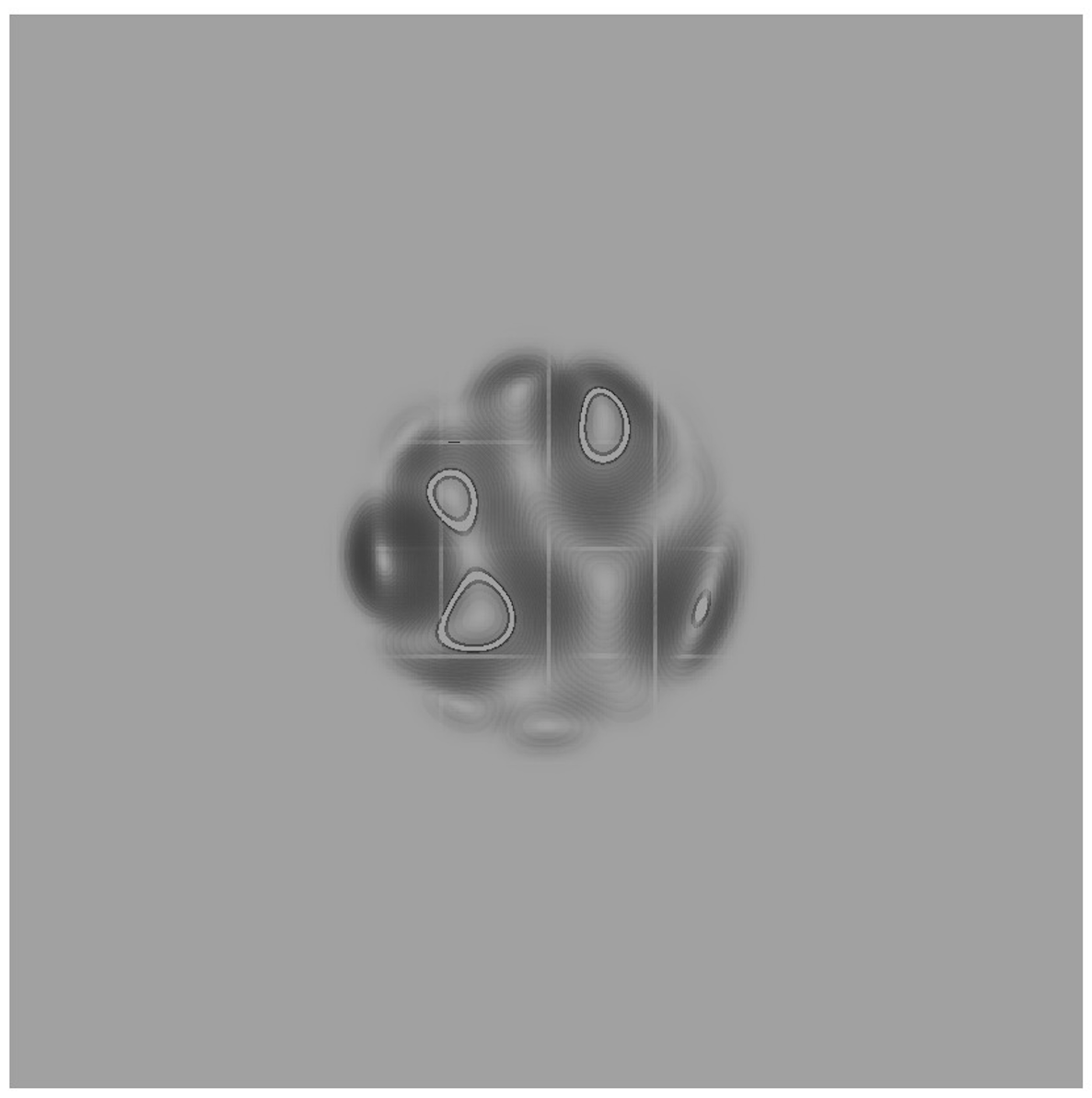}
\caption{Plot of the equimolar plane between the heavy and light fluids, representing the interface.  {\bf Left:} Planar initial conditions form Thornber et al.  {\bf Center:} Present ``curtain'' geometry.  {\bf Right:} Present ``plume'' geometry }
\label{fig:init}
\end{figure}

From Thornber et al (see equation (4) in \cite{thornber.pof.2017}), we introduce a mask function $w(y,z)$ to the volume fraction field, $f_1$, which contains the initial perturbations which can then be written as:
\begin{align}
    f_1(x,y,z) = \frac{1}{2} \text{erfc} \left(  \frac{\sqrt{\pi} [ x-S(y,z)w(y,z) ]    }{ \delta } \right).
\end{align}
The weight function is then given as

\begin{align}
    w(y,z)  = \frac{1}{2}\left( 1 - \tanh\left( \frac{r - r_0 } {\delta_w} \right)   \right)
\end{align}
where $r_0 = 2\pi/6$, $\delta_w = 2\pi/60$, and $\delta = 2\pi/32$ for all cases.  For the ``curtain'' cases, $r=|y-\pi|$ and for the ``plume'' cases, $r=\sqrt{ (y-\pi)^2 + (z-\pi)^2  }$.  From \cite{thornber.pof.2017} planar perturbations are contained in $S(y,z)$ and have a characteristic length scale of $\lambda_0$ and a characteristic growth rate of $\dot{W}$, which is given in \cite{thornber.pof.2017}. The resultant initial interfaces for the unmasked, curtain, and plume geometries are shown in Figure \ref{fig:init}.

%

\section{Results}
\label{sec:results}
\noindent The LES calculations discussed in Section \ref{sec:methods} are performed up to a time of $\tau = 6$, where $\tau=\dot{W}/\lambda_0 t$.  The temporal evolution of the mixing layer growth for the curtain and plume cases can be seen in Figure \ref{fig:curtainHistory} and Figure \ref{fig:plumeHistory}, respectively. 

\begin{figure}[h]
\centering
%
\ffbox{\includegraphics[trim={540pt 55pt 275pt 70pt},clip,width=.115\textwidth]{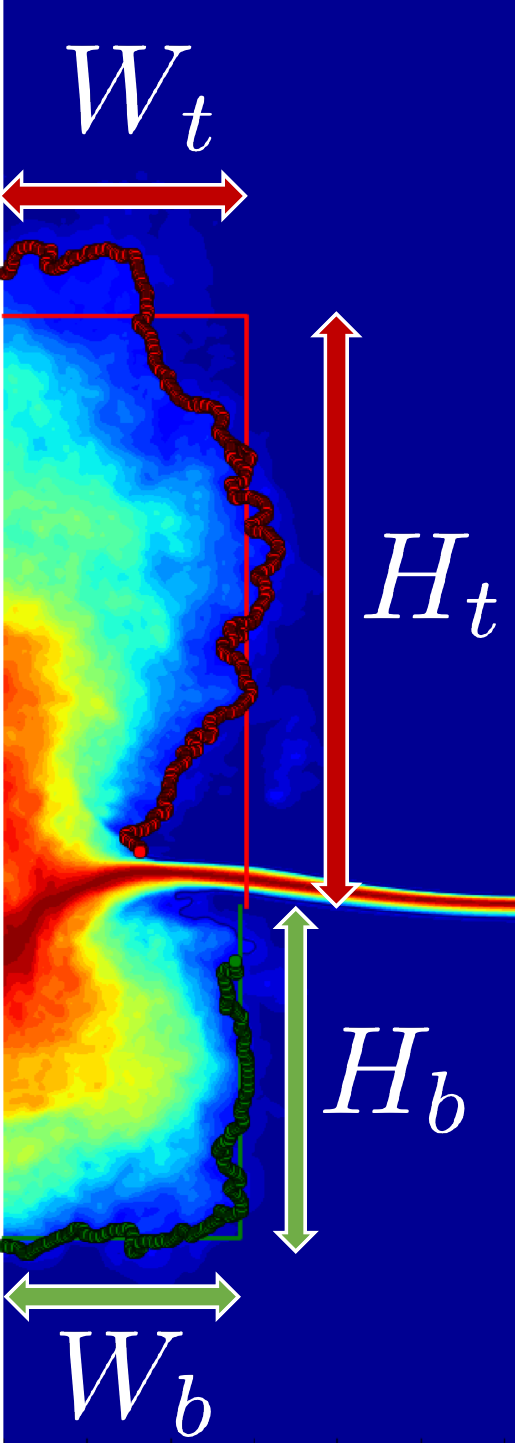}}
\ffbox{\includegraphics[trim={25pt 10pt 50pt 25pt},clip,width=.41\textwidth]{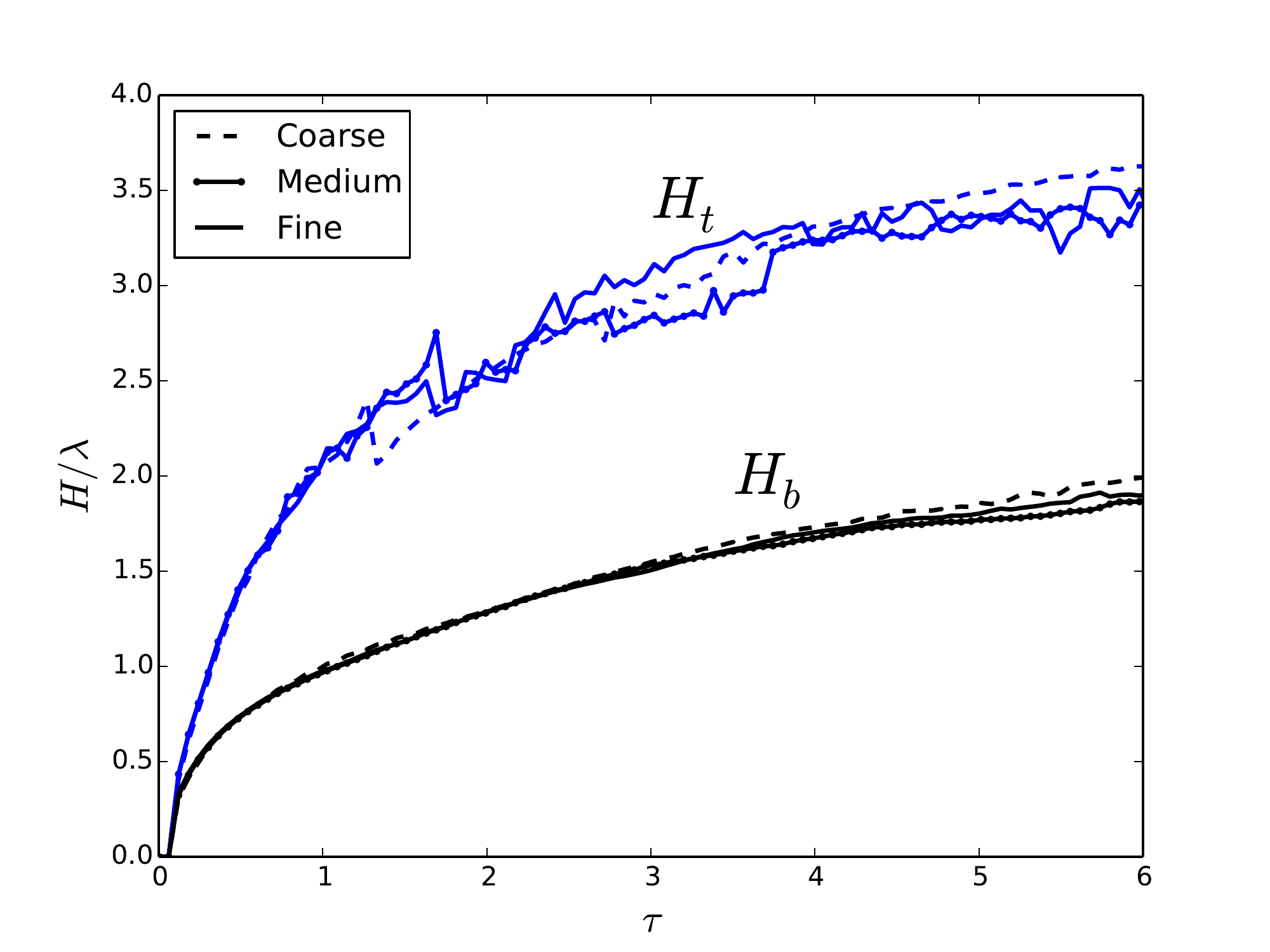}}
\ffbox{\includegraphics[trim={25pt 10pt 50pt 25pt},clip,width=.41\textwidth]{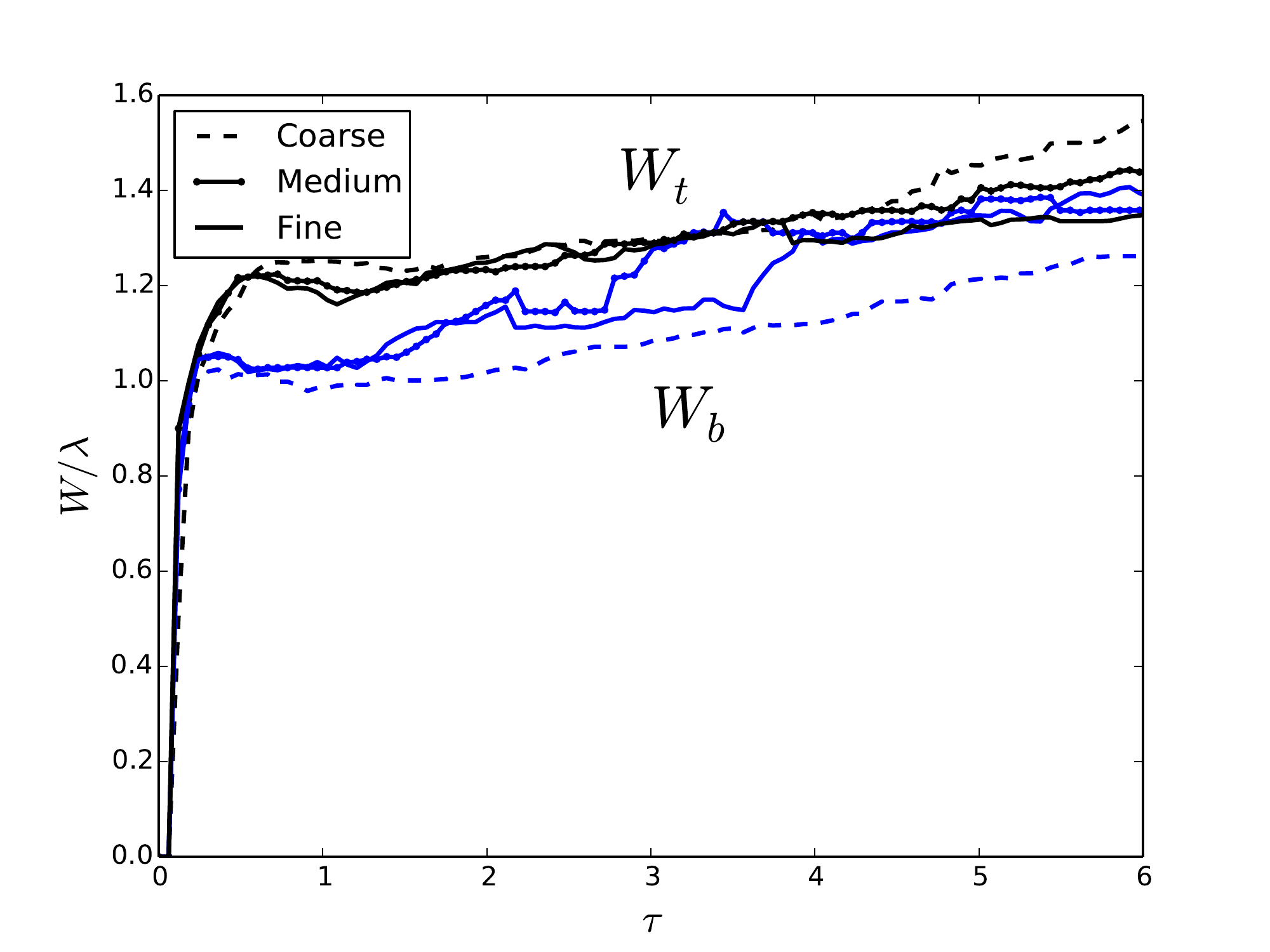}}
\caption{Mixing length scales of the ``curtain'' case in the vertical ($H$) and horizontal ($W$) directions in the top and bottom fluids (subscript $t$ and $b$, respectively).  {\bf Left:} Contour of $4\left< Y_t Y_b \right>$ with bounding box and .1 level contour in red and green {\bf Center:} Mixing heights in the top ($H_t$, blue) and bottom ($H_b$, black) regions for the three mesh resolutions.   {\bf Right:}  Mixing widths in the top ($W_t$, blue) and bottom ($W_b$, black) regions for the three mesh resolutions.   }
\label{fig:curtain_length}
\end{figure}

\begin{figure}[h]
\centering
%
\ffbox{\includegraphics[trim={185pt 55pt 630pt 70pt},clip,width=.115\textwidth]{figure_mod.pdf}}
\ffbox{\includegraphics[trim={25pt 10pt 50pt 25pt},clip,width=.41\textwidth]{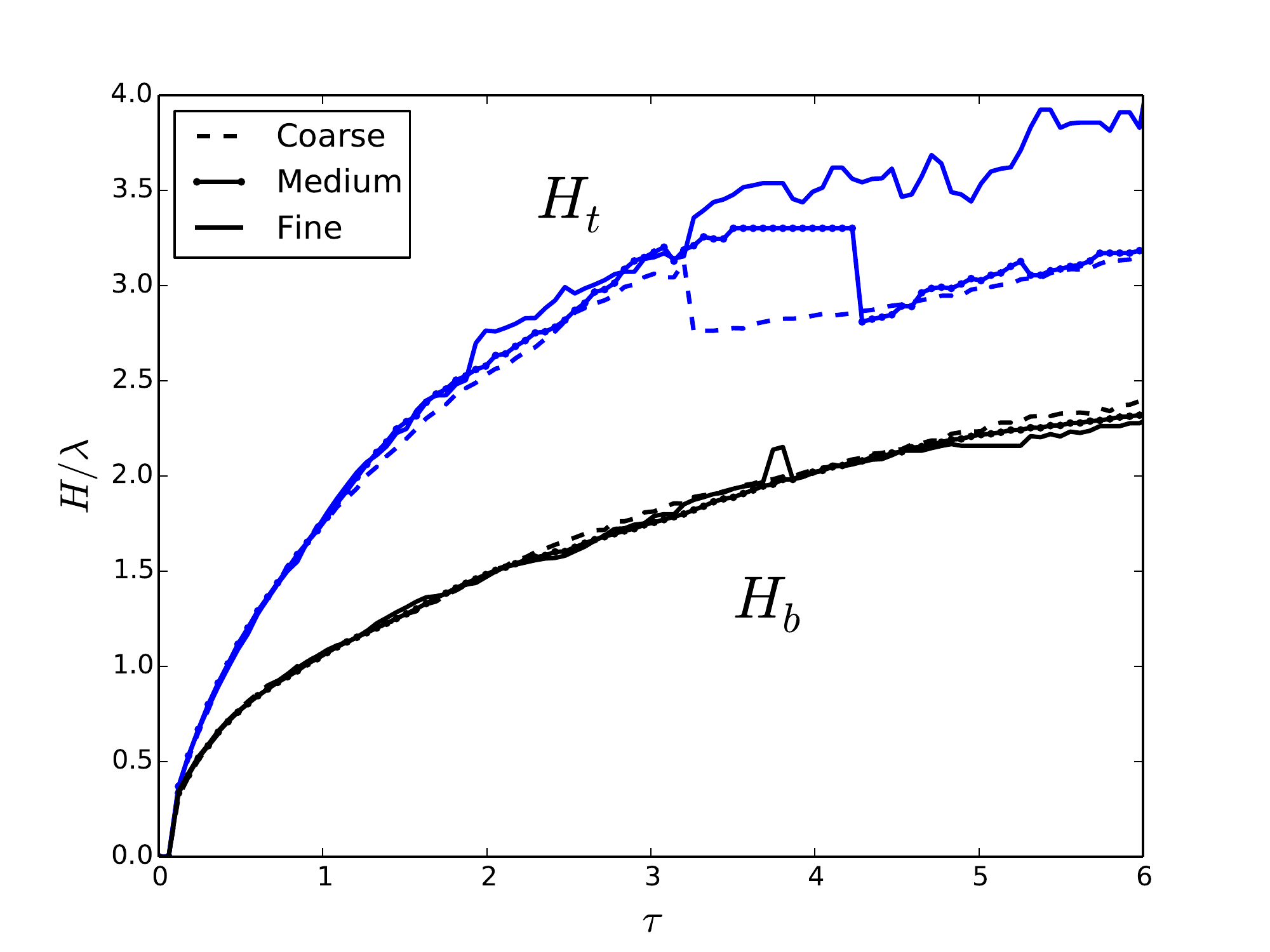}}
\ffbox{\includegraphics[trim={25pt 10pt 50pt 25pt},clip,width=.41\textwidth]{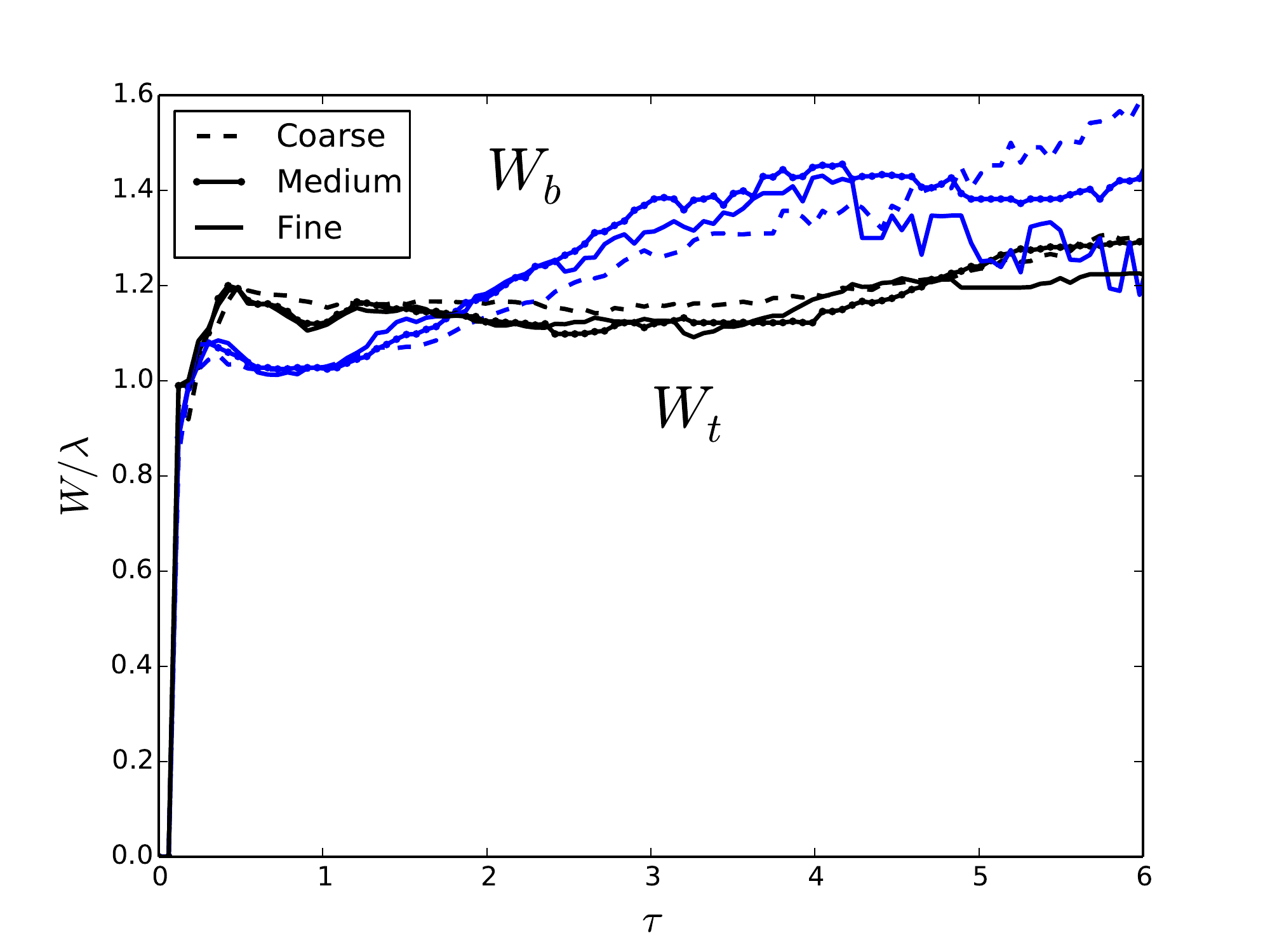}}
\caption{Mixing length scales of the ``plume'' case in the vertical ($H$) and horizontal ($W$) directions in the top and bottom fluids (subscript $t$ and $b$, respectively).  {\bf Left:} Contour of $4\left< Y_t Y_b \right>$ with bounding box and .1 level contour in red and green {\bf Center:} Mixing heights in the top ($H_t$, blue) and bottom ($H_b$, black) regions for the three mesh resolutions.   {\bf Right:}  Mixing widths in the top ($W_t$, blue) and bottom ($W_b$, black) regions for the three mesh resolutions.   }
\label{fig:plume_length}
\end{figure}

The qualitative behavior is very similar between the two cases; small scale perturbations grow to large coherent bubbles and spikes which then lead the transitional turbulent behavior and small scale mixing of the two fluids.  Both cases show a top-bottom asymmetry in the mixing layers as would be expected given the initial Atwood number of the two fluids.  The extent of the vertical (height) and horizontal (width) mixing layer in the top and bottom fluids is quantified by constructing a ``best fit'' bounding box in each fluid around the contour of $4\left< Y_t Y_b \right> = .1$ and taking the length scales from these rectangles, as depicted in Figures \ref{fig:curtain_length} and \ref{fig:plume_length}.

\newlength{\widthSixth}
\setlength{\widthSixth}{.145\textwidth}

\newlength{\xLeft}
\setlength{\xLeft}{220pt}

\begin{figure}[h!]
\centering
\ffbox{\includegraphics[trim={220pt 180pt 850pt 120pt},clip,width=\widthSixth]{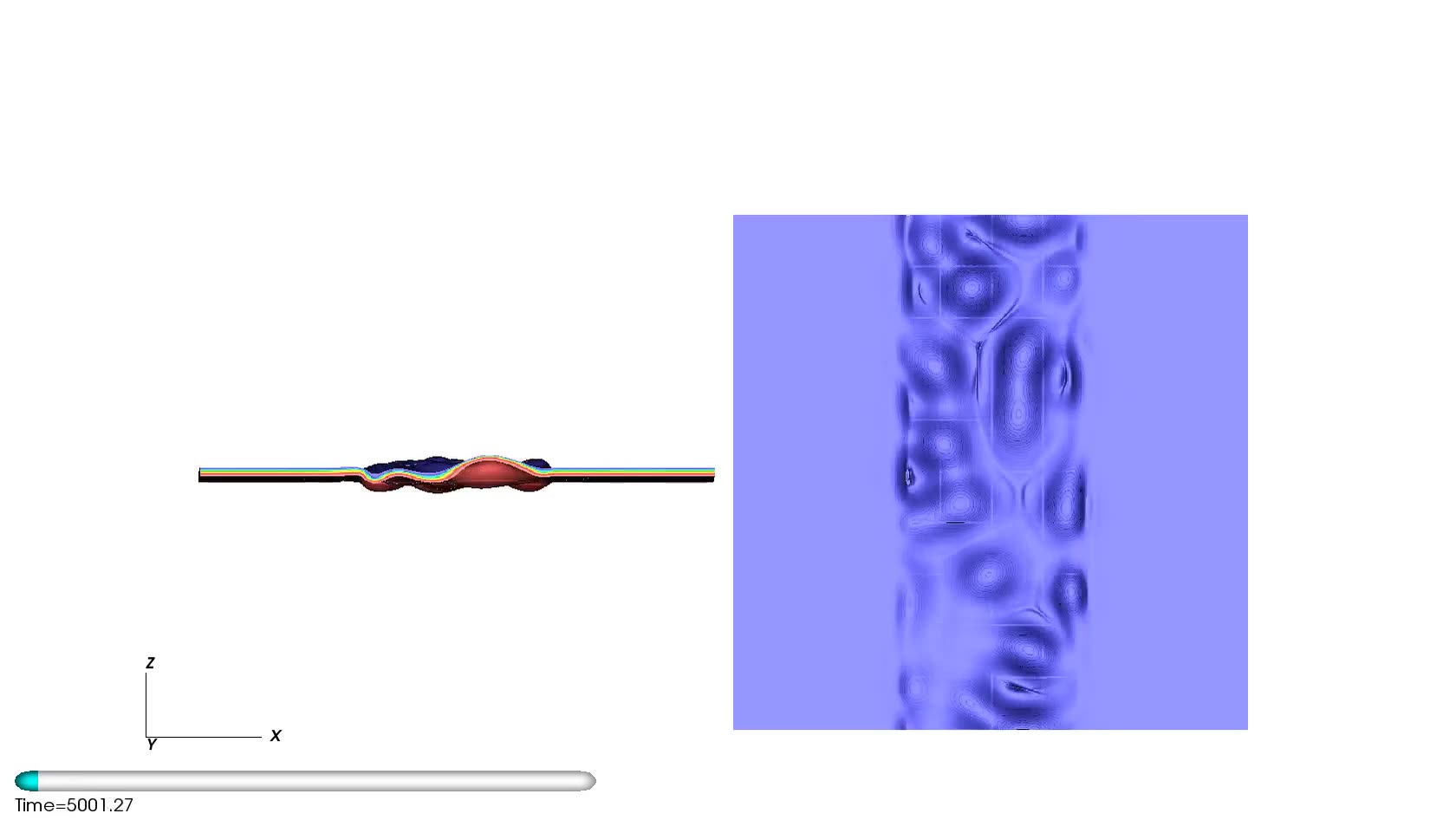}}
\ffbox{\includegraphics[trim={220pt 180pt 850pt 120pt},clip,width=\widthSixth]{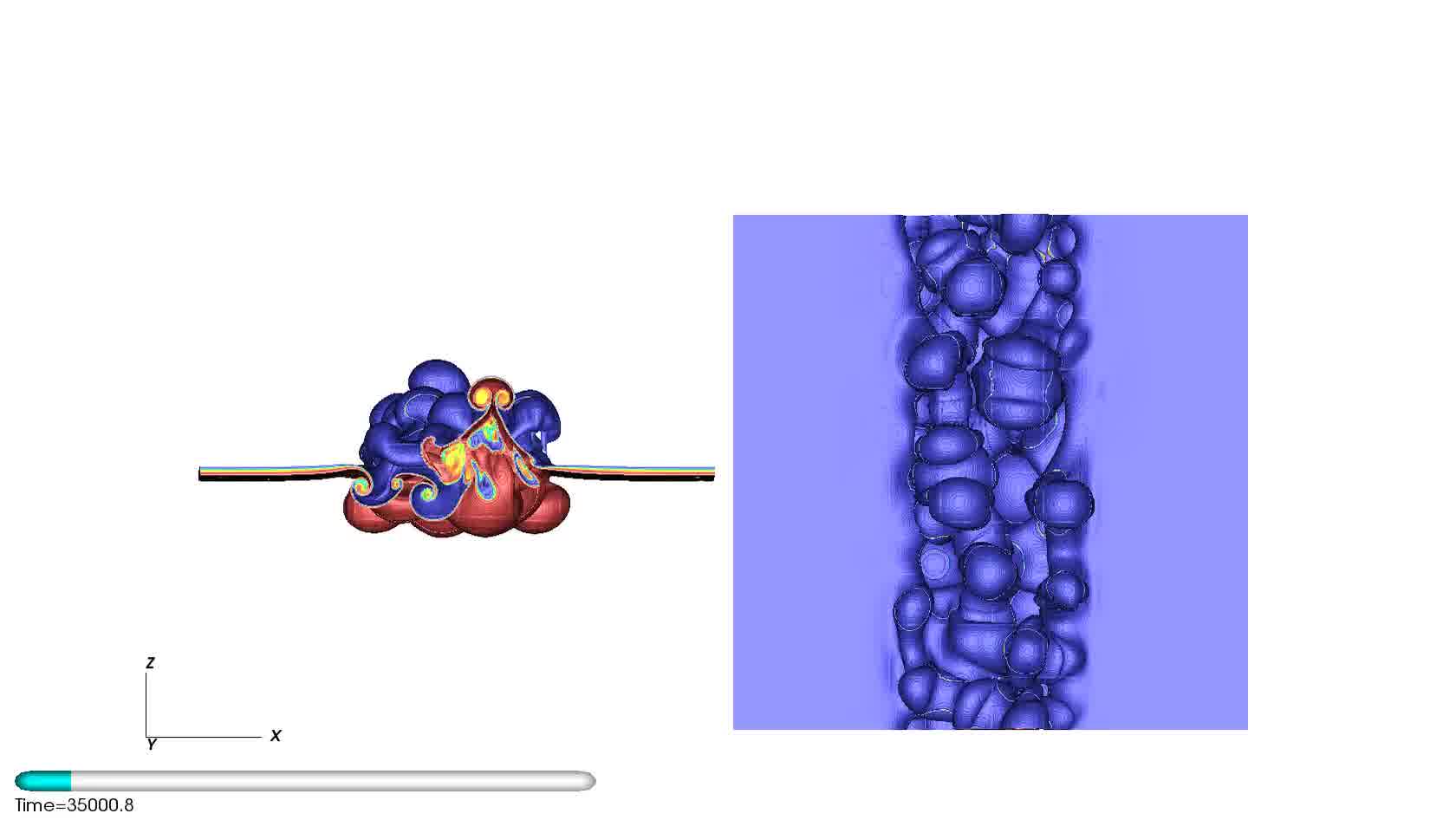}}
\ffbox{\includegraphics[trim={220pt 180pt 850pt 120pt},clip,width=\widthSixth]{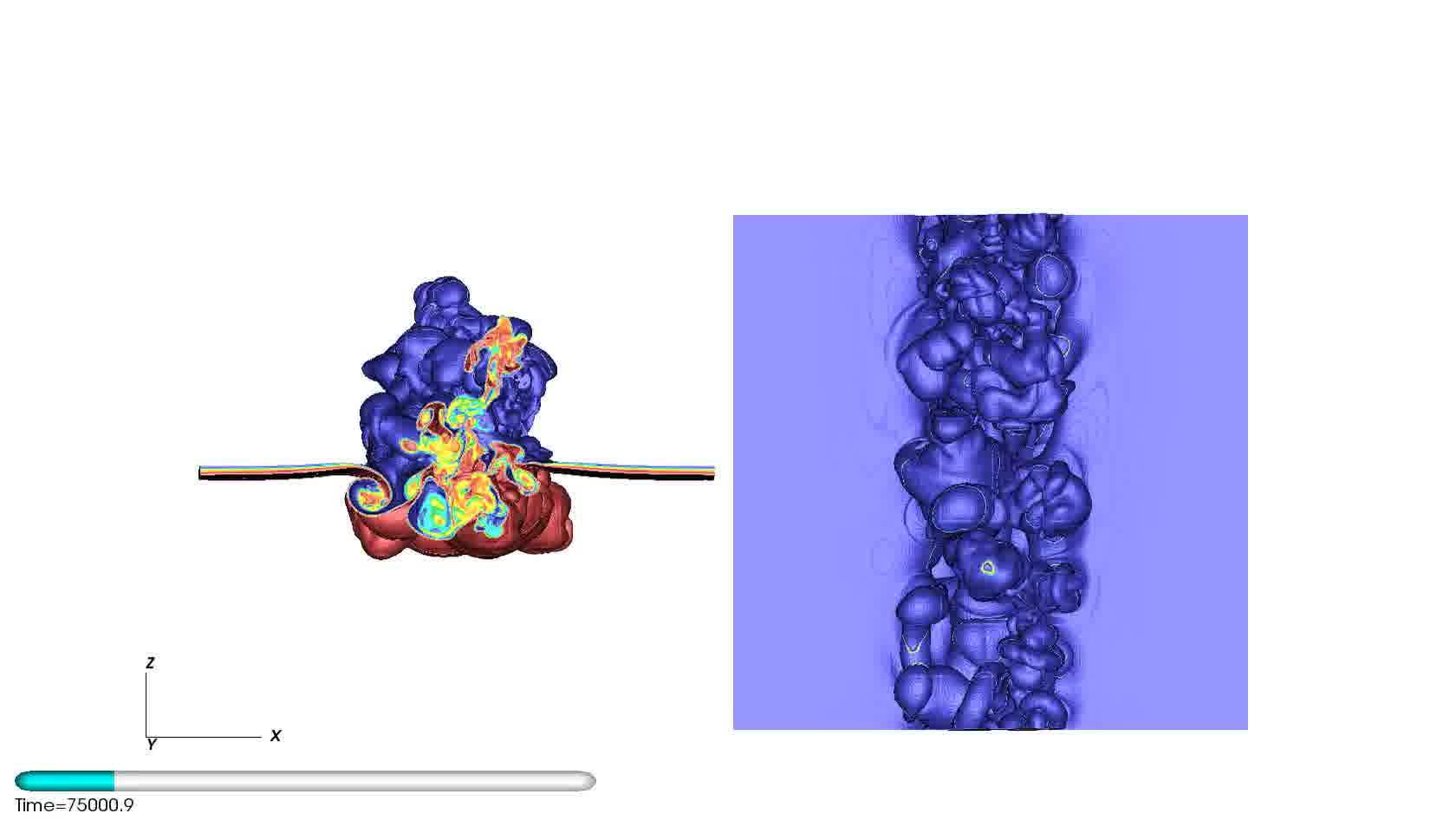}}
\ffbox{\includegraphics[trim={220pt 180pt 850pt 120pt},clip,width=\widthSixth]{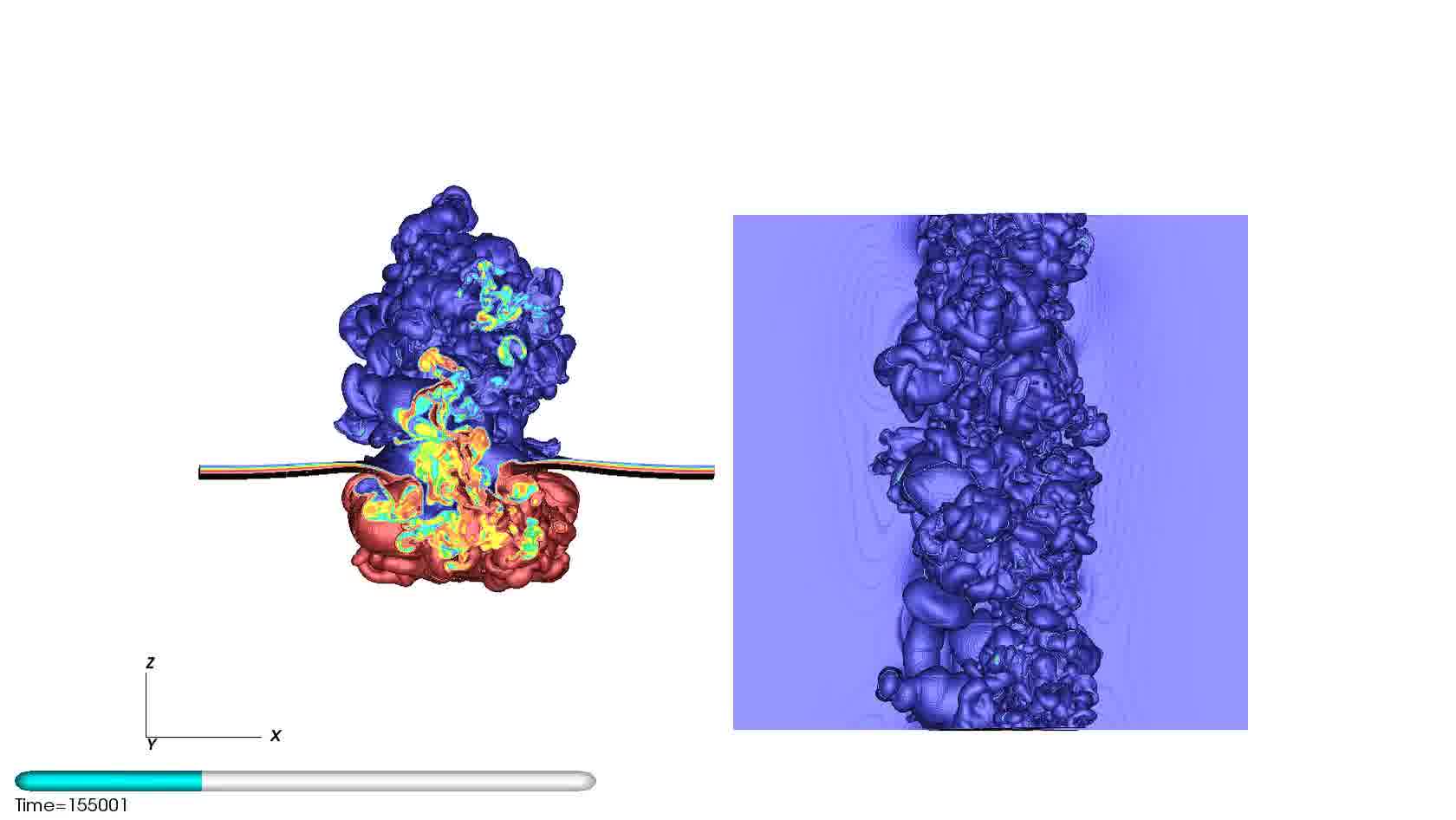}}
\ffbox{\includegraphics[trim={220pt 180pt 850pt 120pt},clip,width=\widthSixth]{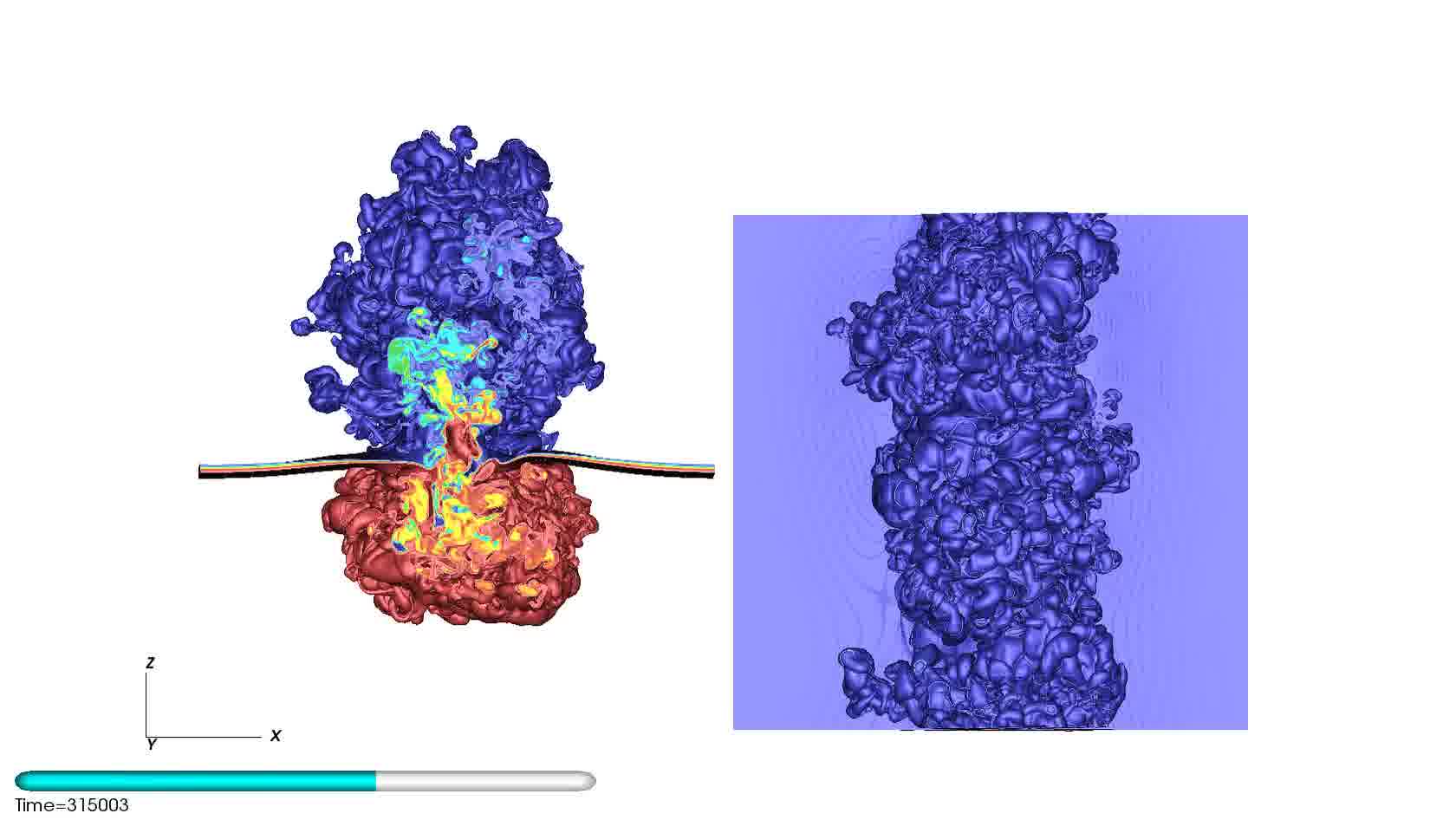}}
\ffbox{\includegraphics[trim={220pt 180pt 850pt 120pt},clip,width=\widthSixth]{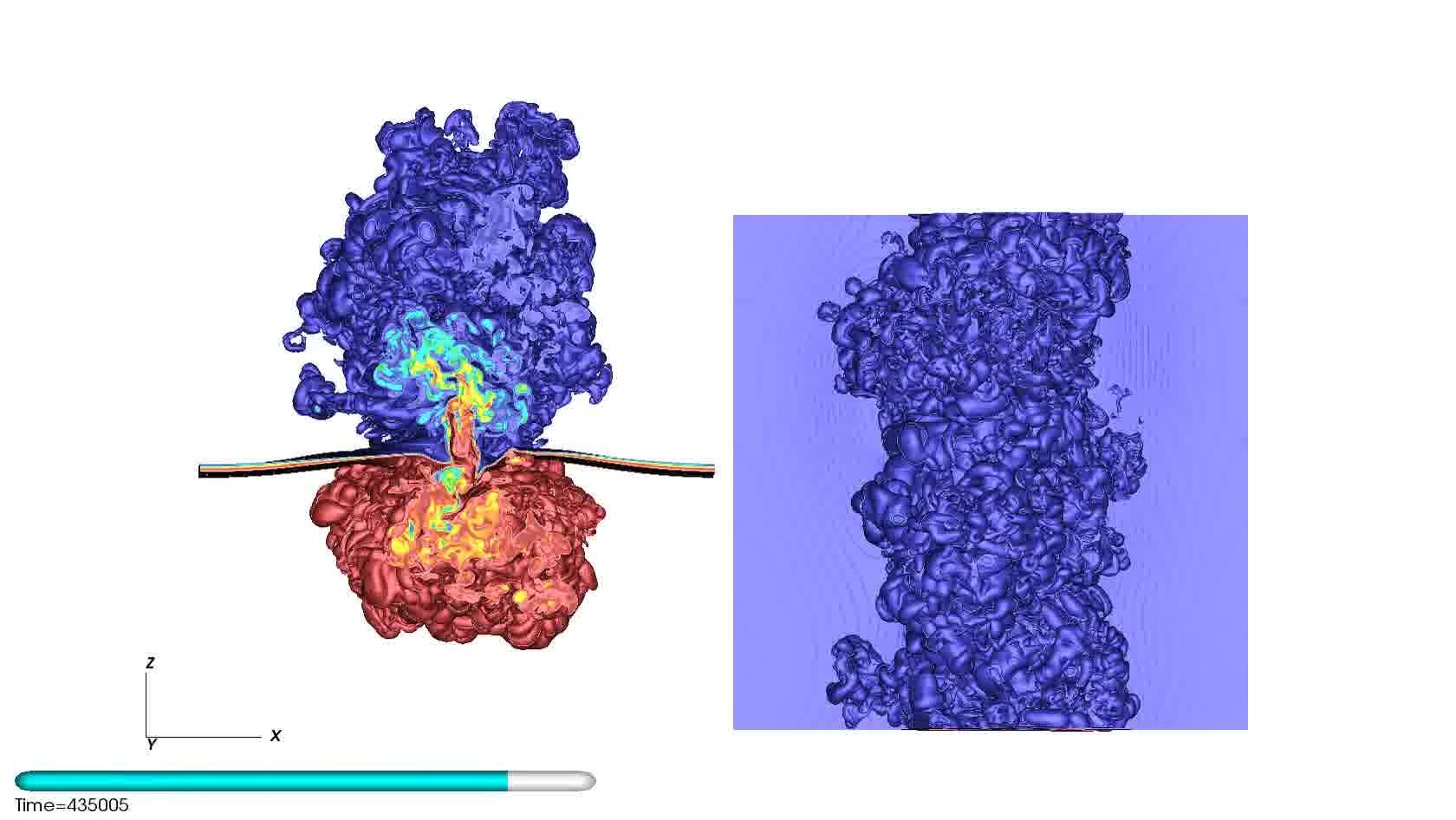}} \\
\ffbox{\includegraphics[trim={840pt 80pt 230pt 220pt},clip,width=\widthSixth]{movie1_0002.jpg}}
\ffbox{\includegraphics[trim={840pt 80pt 230pt 220pt},clip,width=\widthSixth]{movie1_0010.jpg}}
\ffbox{\includegraphics[trim={840pt 80pt 230pt 220pt},clip,width=\widthSixth]{movie1_0020.jpg}}
\ffbox{\includegraphics[trim={840pt 80pt 230pt 220pt},clip,width=\widthSixth]{movie1_0040.jpg}}
\ffbox{\includegraphics[trim={840pt 80pt 230pt 220pt},clip,width=\widthSixth]{movie1_0080.jpg}}
\ffbox{\includegraphics[trim={840pt 80pt 230pt 220pt},clip,width=\widthSixth]{movie1_0110.jpg}} \\
\caption{Time sequence of the mixing layer region for the ``curtain'' case; volume fraction between .1 and .9 of the side-on (top row) and top-down (bottom row) views.  Times are (left to right) $\tau=0.06, 0.42, 0.91, 1.87, 3.80, 5.25$.
 }
\label{fig:curtainHistory}
\end{figure}

\begin{figure}[h!]
\centering
\ffbox{\includegraphics[trim={100pt 180pt 970pt 120pt},clip,width=\widthSixth]{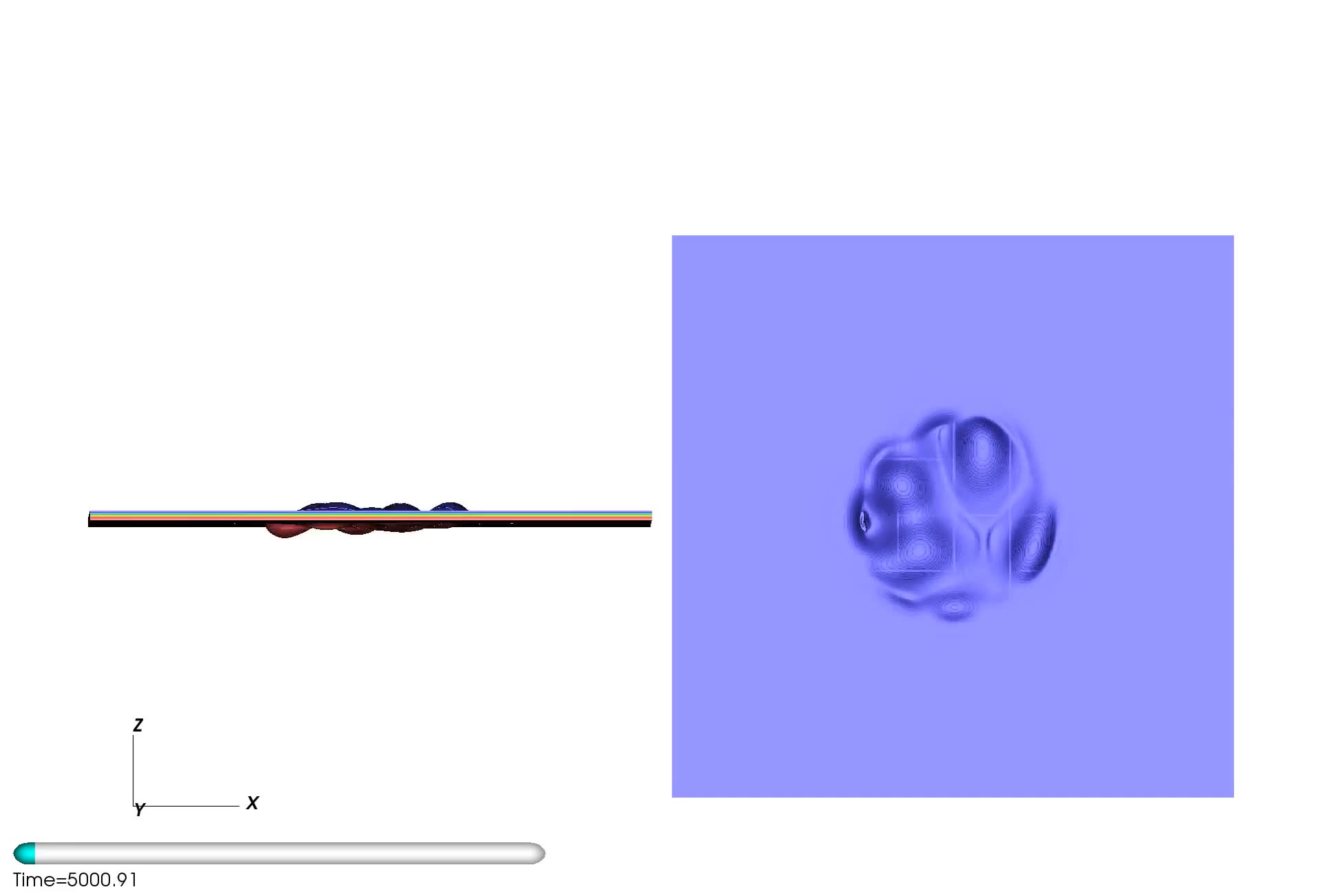}}
\ffbox{\includegraphics[trim={100pt 180pt 970pt 120pt},clip,width=\widthSixth]{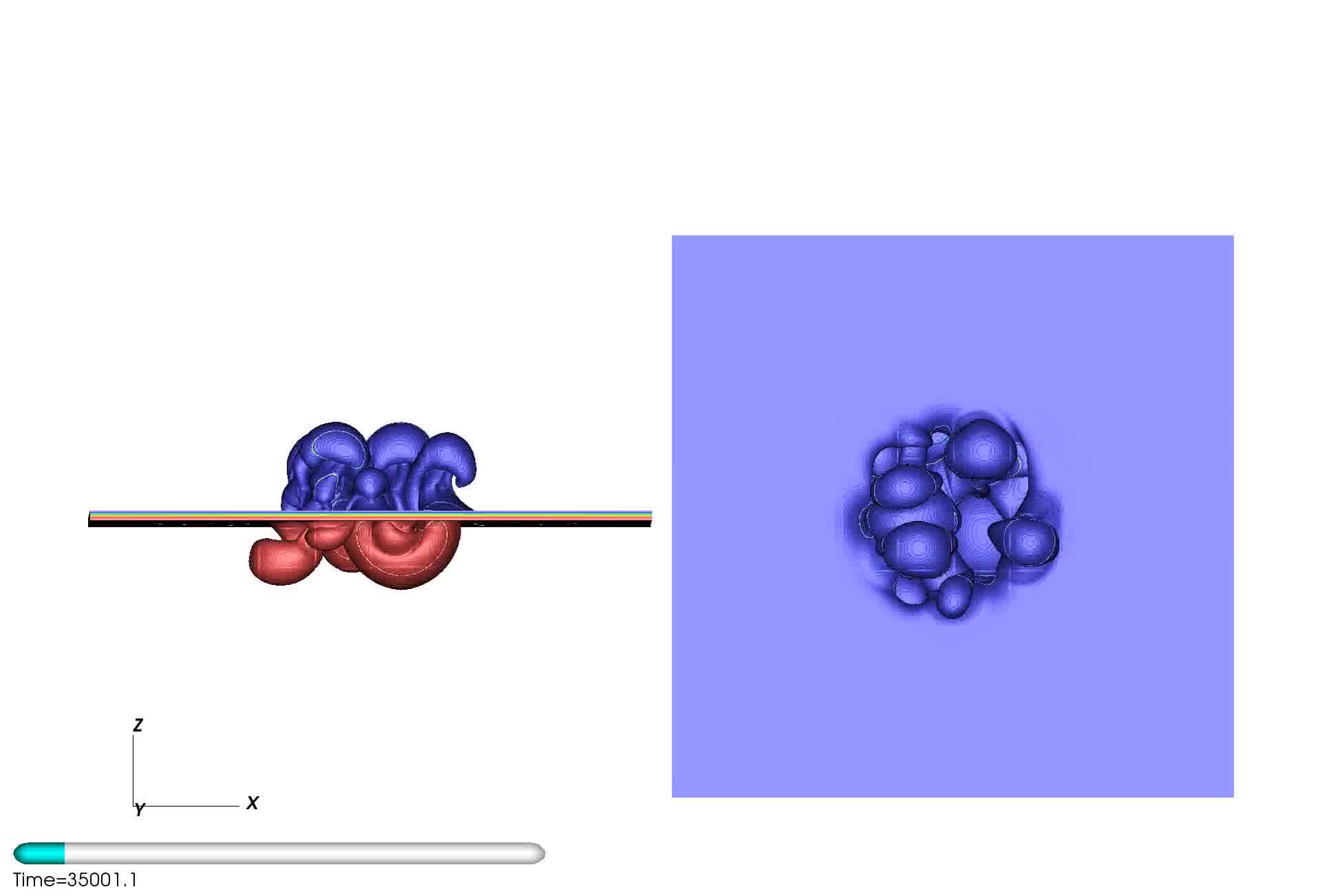}}
\ffbox{\includegraphics[trim={100pt 180pt 970pt 120pt},clip,width=\widthSixth]{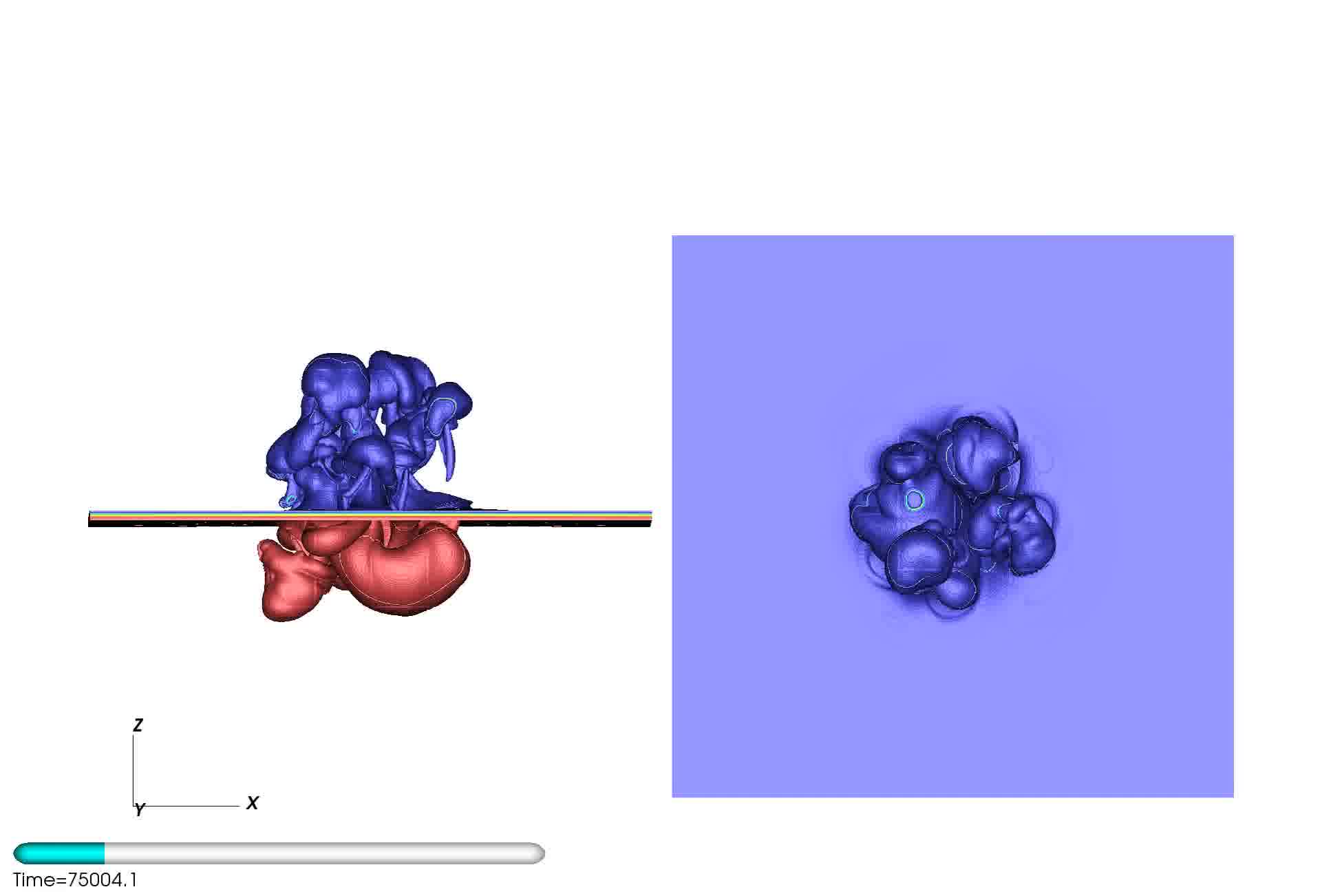}}
\ffbox{\includegraphics[trim={100pt 180pt 970pt 120pt},clip,width=\widthSixth]{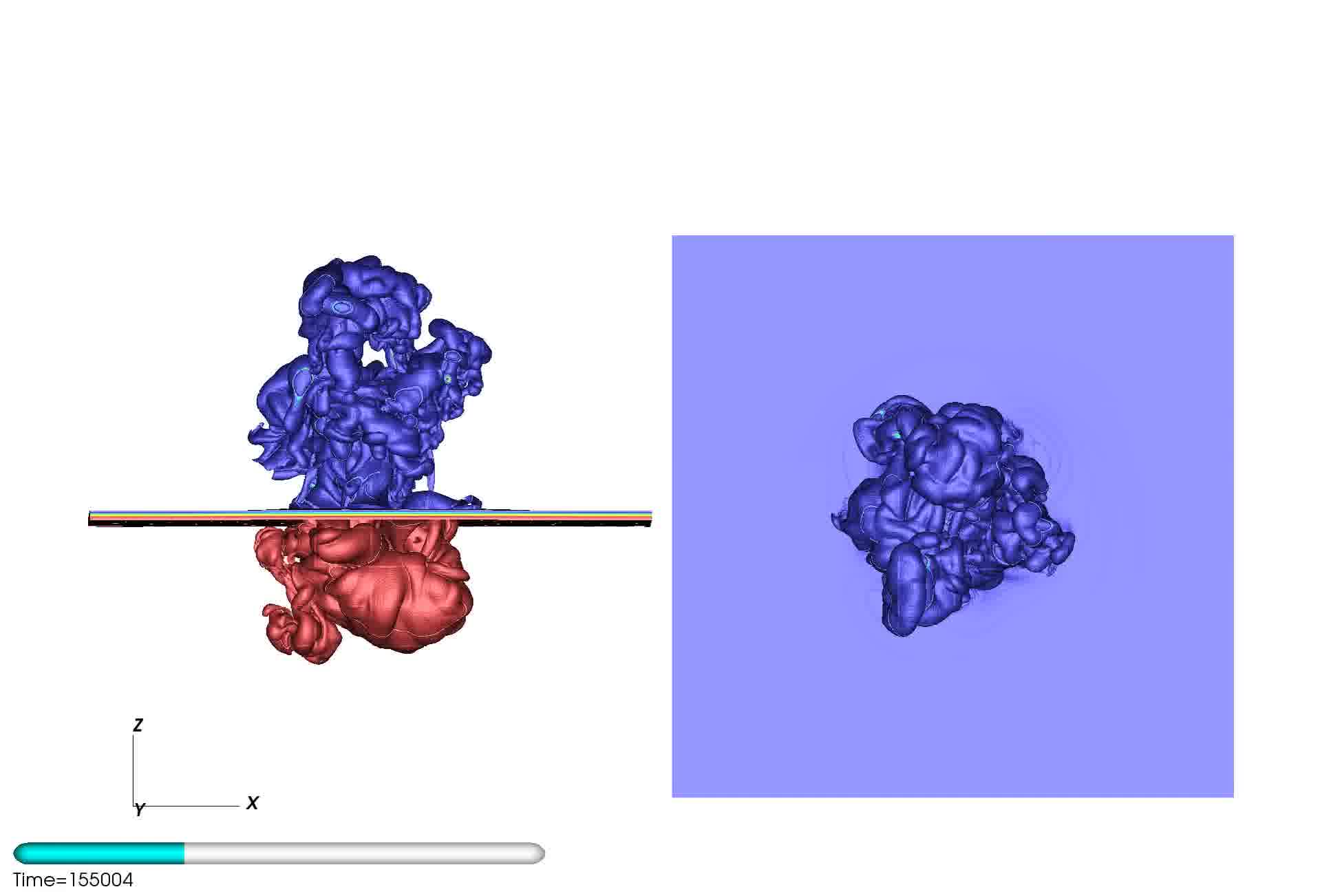}}
\ffbox{\includegraphics[trim={100pt 180pt 970pt 120pt},clip,width=\widthSixth]{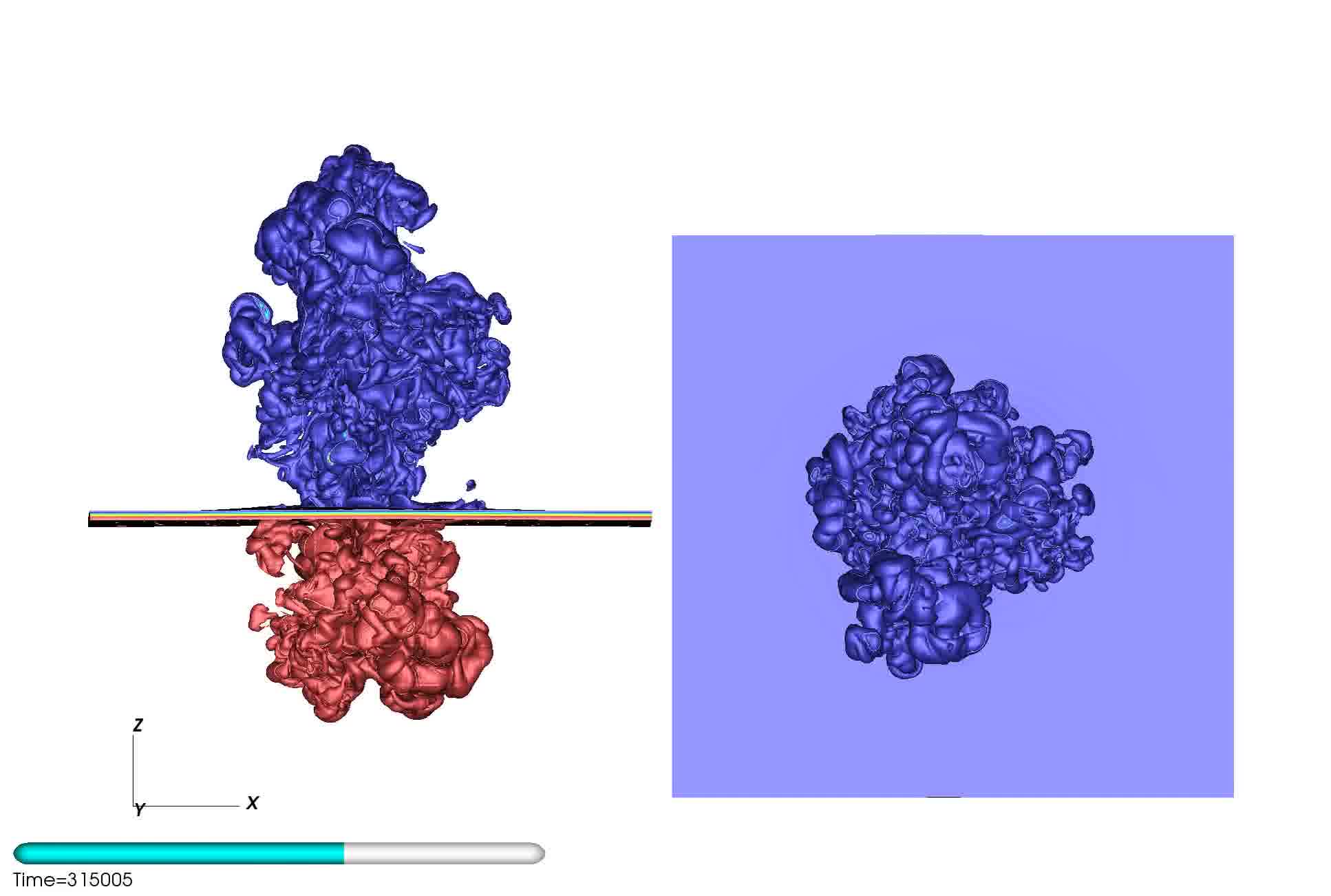}}
\ffbox{\includegraphics[trim={100pt 180pt 970pt 120pt},clip,width=\widthSixth]{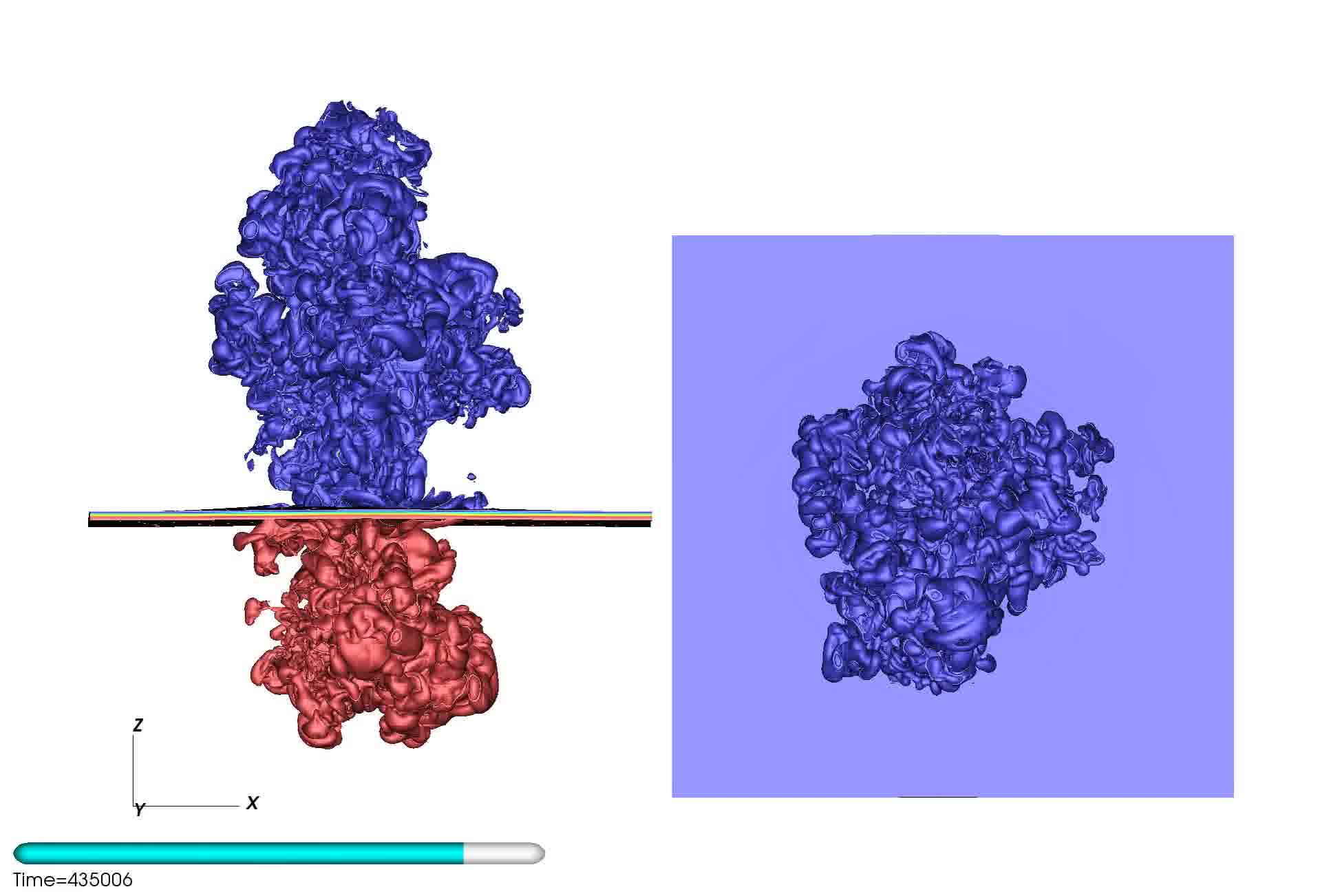}} \\
\ffbox{\includegraphics[trim={990pt 80pt 140pt 220pt},clip,width=\widthSixth]{movie2_0002.jpg}}
\ffbox{\includegraphics[trim={990pt 80pt 140pt 220pt},clip,width=\widthSixth]{movie2_0010.jpg}}
\ffbox{\includegraphics[trim={990pt 80pt 140pt 220pt},clip,width=\widthSixth]{movie2_0020.jpg}}
\ffbox{\includegraphics[trim={990pt 80pt 140pt 220pt},clip,width=\widthSixth]{movie2_0040.jpg}}
\ffbox{\includegraphics[trim={990pt 80pt 140pt 220pt},clip,width=\widthSixth]{movie2_0080.jpg}}
\ffbox{\includegraphics[trim={990pt 80pt 140pt 220pt},clip,width=\widthSixth]{movie2_0110.jpg}} \\
\caption{Time sequence of the mixing layer region for the ``plume'' case; volume fraction between .1 and .9 of the side-on (top row) and top-down (bottom row) views.  Times are (left to right) $\tau=0.06, 0.42, 0.91, 1.87, 3.80, 5.25$.
 }
\label{fig:plumeHistory}
\end{figure}

Figures \ref{fig:curtain_length} and \ref{fig:plume_length} show that the mixing length scales (between the three mesh resolutions) vary by less than 3\% for $\tau < 3$.  Larger variations later in time (especially for the ``plume'' case) appeared to be due to edge features of the mixing layer, to which the contouring algorithm was sensitive.

The mixing heights in the top and bottom fluids ($H_t,H_b$) appear to be following a $t^\theta$ like growth rate \cite{zhou.pr.2017-1,zhou.pr.2017-2}. 
The mixing widths ($W_t,W_b$), however, appear to be following a totally different growth rate behavior and very little growth occurs in the lateral direction.  The lack of growth creates a mixing layer with a growing aspect ratio that is clearly visualized in Figures \ref{fig:curtainHistory} and \ref{fig:plumeHistory}.

\begin{figure}[h]
\centering
\ffbox{
\begin{tikzpicture}
    \draw (0, 0) node[inner sep=0] {\includegraphics[trim={155pt 90pt 850pt 90pt},clip,width=.15\textwidth]{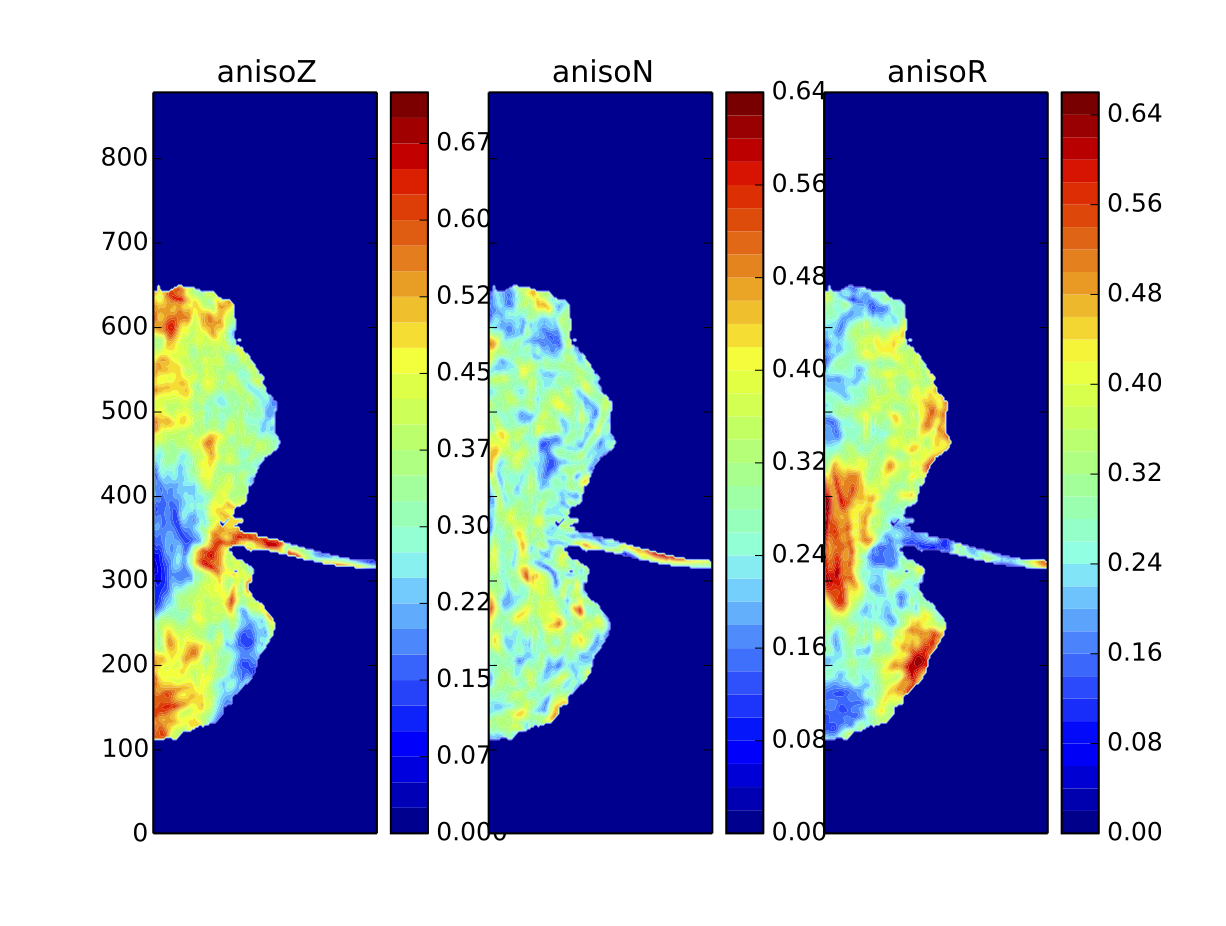}};
    \draw (0.5,-3) [color=white] node { \Large$ \frac{\left<u_r^`u_r^`\right>}{  \left<u_i^`u_i^`\right> }$};
    \draw (0,-4.2) [color=black] node { (a)};    
    \end{tikzpicture}
}
\ffbox{
\begin{tikzpicture}
    \draw (0, 0) node[inner sep=0] {\includegraphics[trim={828pt 90pt 177pt 90pt},clip,width=.15\textwidth]{image101.jpeg}};
    \draw (0.5,-3) [color=white] node { \Large$ \frac{\left<u_z^`u_z^`\right>}{  \left<u_i^`u_i^`\right> }$};
    \draw (0,-4.2) [color=black] node { (b)};    
    \end{tikzpicture}
}
\ffbox{
\begin{tikzpicture}
    \draw (0, 0) node[inner sep=0] {\includegraphics[trim={53pt 90pt 50pt 85pt},clip,width=.166\textwidth,height= .498\textwidth]{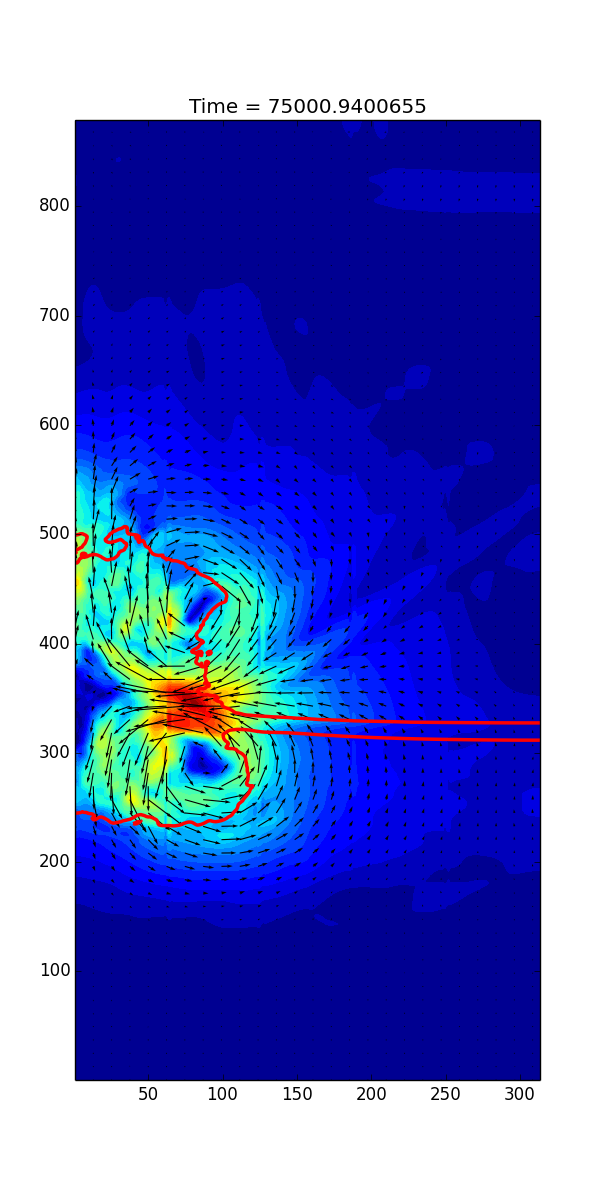}};
    \draw (0,-3) [color=white] node {\Large $|\vec{u}|$};
    \draw (0,-4.2) [color=black] node { (c)};
\end{tikzpicture}
}
\ffbox{
\begin{tikzpicture}
    \draw (0, 0) node[inner sep=0] {\includegraphics[trim={53pt 90pt 50pt 85pt},clip,width=.166\textwidth,height= .498\textwidth]{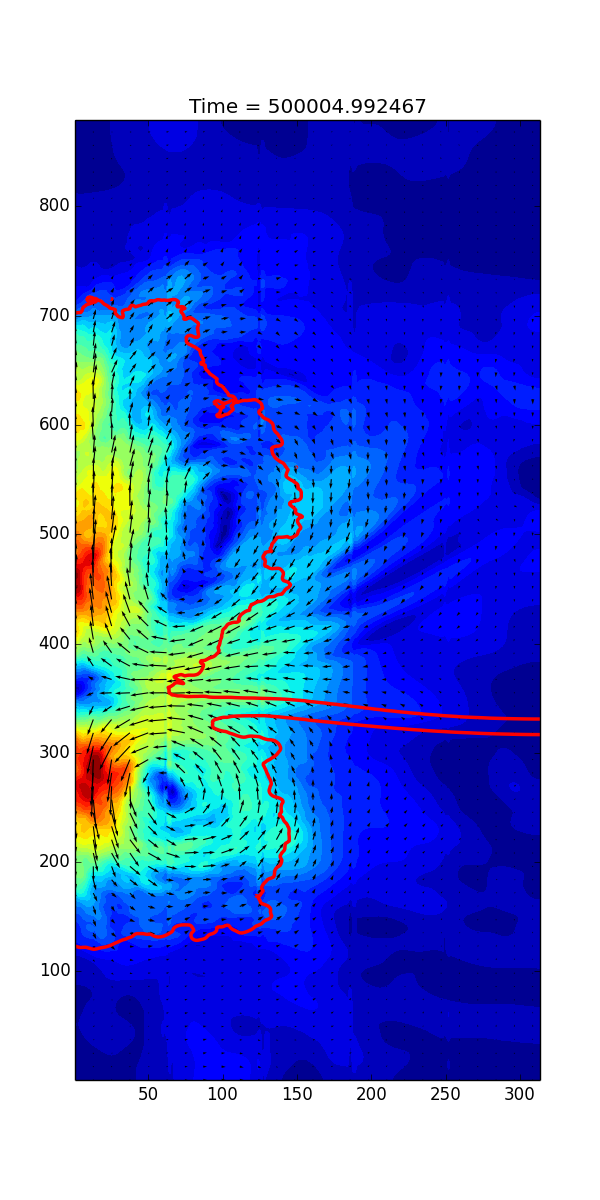}};
    \draw (0,-3) [color=white] node {\Large $|\vec{u}|$};
    \draw (0,-4.2) [color=black] node { (d)};
\end{tikzpicture}
}
\caption{Contours of Reynolds stress anisotropy in the radial (a) and vertical (b) directions at $\tau=6$.  Anisotropy follows the local velocity maxima created by the vortex pair which forms as the layer transitions to turbulence at $\tau=2$ (c) and persists through late time, $\tau=6$ (d). }
\label{fig:aniso}
\end{figure}

Given that RM is a purely decaying process, its somewhat surprising that the mixing layer maintains and grows in spatial inhomogeneity with time that is not present in the standard planar case.  
The flow would be expected to relax to statistical self-similarity once the scale of the plumes substantially exceeds the lateral scale of the initial patch, but this condition has not been reached in the calculations presented here.
To explore the mixing scales anisotropy, the Reynolds stress is computed in the principal directions, and normalized by its sum over all directions.  This anisotropy is shown in Figure \ref{fig:aniso}(a,b).  

Local maxima of the radial Reynolds stress anisotropy, $\left< u'_r u'_r \right> / \left< u'_i u'_i \right>$, in Figure \ref{fig:aniso}(a) are localized near the mixing region front and the initial interface location.  Local maxima of the veritical Reynolds stress anisotropy, $\left< u'_z u'_z \right> / \left< u'_i u'_i \right>$, in Figure \ref{fig:aniso}(b) are localized near the core of mixing region.  The $\left< \right>$ operator denotes $xr$-planar averages via a binning operation.   

Inspection of the mean velocity magnitude reveals the source of this anisotropy as a vortex pair which which forms early on ($\tau < 2$) during the instability transition process and persists to late time.  Figure \ref{fig:aniso}(c) and \ref{fig:aniso}(b) show contours of the mean velocity magnitude with velocity vectors and the mixing layer edge drawn in black and red, respectively.  The persistent vortex entrains pure fluid into the mixing along the interface, creating local variations in the Reynolds stress anisotropy and supressing the lateral spreading of the mixing layer.

%
%

\section{Discussion}
\label{sec:discussion}
%
%

\noindent The results presented here indicate that significant multi-dimensional effects are present in both the curtain and plume geometries.  The mixing layer growth, anisotropy, and mean flow are independent of the mesh resolutions to within 5 and 15\% for the curtain and plume geometries, respectively. The multi-dimensional effects appear during the transition process of the instability and persist as large scale vortices located at the edges of the mixing layer.  

Preliminary studies (the results of which are beyond the scope of the present work and not shown here) using a simple K-L RANS \cite{morgan.shockwaves.2016} model to capture the patch of instability growth fail to capture turbulent transition and therefore don't accurately predict the two-dimensional mean flow.  The vertical and horizontal mixing length scales are under-predicted and over-predicted, respectively, by a factor of two.  The vortex pair captured by the LES mean flow solution is absent in the RANS calculation.  A modification to the RANS modeling approach and/or directly capturing the mean flow would be required to capture these complex mixing layers accurately.  Subsequent work is planned to assess modeling approaches.
This work was performed under the auspices of the U.S. Department of Energy by Lawrence Livermore National Laboratory under Contract DE-AC52-07NA27344.




\end{document}